\documentclass[prb,twocolumn,showpacs,floatfix,preprintnumbers]{revtex4}

\usepackage{amsmath,amssymb,amsfonts}
\usepackage{bm,graphicx,color}

\newcommand{\beq}{\begin{equation}}
\newcommand{\eeq}{\end{equation}}
\newcommand{\beqa}{\begin{eqnarray}}
\newcommand{\eeqa}{\end{eqnarray}}
\newcommand{\beal}{\begin{align}}
\newcommand{\eal}{\end{align}}

\newcommand{\BK}{{\bm k}}

\begin{document}

  \title{Correlations in a band insulator}

  \author{Michael Sentef$^1$}
  \email{sentefmi@physik.uni-augsburg.de}
  \author{Jan Kune\v{s}$^1$}
  \author{Philipp Werner$^2$}
  \author{Arno P. Kampf$^1$}
  \affiliation{$^1$Theoretical Physics III, Center for Electronic
          Correlations and Magnetism, Institute of Physics, 
          University of Augsburg, D-86135 Augsburg, Germany \\
          $^2$Theoretical Physics, ETH Zurich, CH-8093 Zurich,
          Switzerland}

  \date{\today}

  \begin{abstract}
    We study a model of a covalent band insulator with on-site Coulomb
    repulsion at half-filling using dynamical mean-field theory.    
    Upon
    increasing the interaction strength the system undergoes a
    discontinuous transition from a correlated band insulator to a Mott
    insulator with hysteretic behavior at low temperatures.
    Increasing the temperature in the band insulator close to the insulator-insulator transition we find a
    crossover to a Mott insulator at elevated temperatures. Remarkably, correlations decrease the energy gap in the correlated band insulator. The gap renormalization can be traced to the low-frequency behavior of the self-energy, analogously to the quasiparticle renormalization in a Fermi liquid. While the uncorrelated band insulator is characterized by a single gap for both charge and spin excitations, the spin gap is smaller than the charge gap in the correlated system. %
  \end{abstract}

  \pacs{71.10.Fd, 71.10.Hf, 71.27.+a%
    \vspace*{-3mm}%
  }

  \maketitle

\section{Introduction}
The role of electron-electron (e-e) interactions in solids is one of the central problems of condensed matter physics.
The Hubbard model with local e-e interaction has become a paradigm for the description of electronic correlations in narrow-band materials. It has been used to investigate electronic correlations in metals and to study the correlation-driven metal-insulator transition.\cite{Imada98} Much less attention has been paid to electronic correlations in band insulators (BI), since the lack of low-energy excitations rendered them less interesting.

However, the discoveries of the quantum Hall effect and Kondo insulators showed that BIs are far from trivial, and recent progress in the topological classification of BIs \cite{Kane05} demonstrates that our understanding of the insulating state is indeed incomplete.
The quest for materials with topologically non-trivial electronic structures suggests to explore heavier elements with strong spin-orbit coupling 
involving $d$ or $f$ electrons \cite{Shitade08} and raises the question about the role of electronic correlations.
The common feature of these materials is that the constituting atoms have partially
filled shells and the gap -- a hybridization gap -- opens due to a particular pattern of inter-atomic
hopping integrals. It has been proposed that similar characteristics apply to
materials such as FeSi, FeSb$_2$ or CoTiSb,\cite{Kunes08} some of which exhibit
strongly temperature dependent magnetic and transport properties reminiscent of Kondo insulators. We call this class of BIs covalent insulators (CI).

Recently the evolution of a band insulator (BI) into a Mott insulator upon increasing the interaction strength
has been studied in the context of the ionic Hubbard model,\cite{Kampf03,Noack04,Aligia04,Garg06,Kancharla07a,Paris07,Craco08,
Byczuk09} a two-band Hubbard model with crystal-field splitting,\cite{Werner07} and a bilayer model with two identical 
Hubbard planes coupled by single-particle hopping.\cite{Moeller99,Fuhrmann06,Kancharla07b,Fabrizio07,Hafermann09} Different scenarios have emerged with the possibility of an intervening phase and continuous or discontinuous transitions at critical interaction strengths.

In order to study the properties of CIs with local e-e interaction we employ the dynamical mean-field theory (DMFT).\cite{Metzner89,MuellerHartmann89,Jarrell92,Georges92,Georges96} 
In particular we are interested in the nature of the interaction-driven transition from a covalent to a Mott insulator, the possible existence
of an intervening metallic phase, the evolution of charge and spin gaps and
the single-particle self-energy as a function of the interaction strength $U$. 

This paper is organized as follows: In Sec.\ \ref{sec:model} we define the model and the methods chosen to study correlations in the covalent band insulator. The subsequent investigation is guided by the following questions: (i) How does the band insulator at weak coupling evolve into the Mott insulator at strong coupling (Sec.\ \ref{sec:3a})? (ii) What is the effect of correlations on the spectral function, and what happens when the temperature is increased in the correlated system (Sec.\ \ref{sec:3b})? (iii) Can we quantify correlation effects in a band insulator by means of concepts analogous to Fermi liquid theory (Sec.\ \ref{sec:3c})? (iv) Regarding the characterization of a simple band insulator as an insulator with identical charge and spin excitation gaps: Is this picture modified by correlations (Sec.\ \ref{sec:3d})? Our results are summarized and conclusions are drawn in Sec.\ \ref{sec:conclusion}. 
%%%%%%%%%%%%%%%%%%%%%%%%%%%%%%%%%%%%%%%%%
\section{Model and methods} %
\label{sec:model}
As a covalent insulator we denote a band insulator with partially 
filled identical local orbitals. This definition implies that the 
band gap is a hybridization gap arising from a particular pattern 
of hopping integrals.
Realizations of the covalent insulator include dimerized or bilayer lattices,\cite{Moeller99,Fuhrmann06,Kancharla07b,Fabrizio07,Hafermann09}
quantum Hall systems with filled Landau levels
or Haldane's model \cite{Haldane88} and the related model of Kane and Mele \cite{Kane05} describing electrons on a honeycomb lattice with broken time-reversal invariance.
We use a simple particle-hole symmetric model at half-filling described by the Hamiltonian
\beqa
H \!&=&\! \sum_{\BK \sigma} 
\left(
\begin{array}{cc}
a_{\BK\sigma}^{\dagger}, &
b_{\BK\sigma}^{\dagger} \\
\end{array}
\right)
{\bf H}({\BK})
\left(
\begin{array}{cc}
a_{\BK\sigma}^{} \\
b_{\BK\sigma}^{} \\
\end{array}
\right)
\nonumber \\
&&\!
+\;U\sum_{i\alpha}n_{i\uparrow\alpha}n_{i\downarrow\alpha} ,
\\
{\bf H}({\BK}) &=& 
\left(
\begin{array}{cc}
\epsilon_\BK & V \\
V & - \epsilon_\BK \\
\end{array}
\right),
\label{eq:hamilt}
\eeqa
with two semi-circular electronic bands of widths $4t$ ($t=1$ in the following) and dispersions $\epsilon_\BK$ and $-\epsilon_\BK$, respectively, corresponding to two sublattices coupled by the $\bm{k}$-independent hybridization $V$ and a local e-e interaction of strength $U$. Here $n_{i\sigma\alpha}=\alpha_{i\sigma}^{\dagger}\alpha_{i\sigma}^{}$ measures the number of electrons with spin $\sigma=\uparrow,\downarrow$ on site $i$ of sublattice $\alpha=a,b$.

We use the DMFT approximation to calculate the local single-particle propagator and the local spin susceptibility, quantities 
which within DMFT depend on the lattice only through the non-interacting density of states and thus are independent of a particular realization of the CI.
The single-particle self-energy $\Sigma(\omega)$ obtained by DMFT is local and fulfils the equations
\beqa
G({\rm i}\omega_n){\bf I} \hspace{-2mm} &=& \hspace{-2mm}
\displaystyle \sum_{\BK}  
\left(
\left({\rm i}\omega_n+\mu-\Sigma({\rm i}\omega_n)\right){\bf I}
-
\hspace{-1mm}
{\bf H}({\BK})
\right)^{-1},\nonumber
\\
G_0^{-1}({\rm i}\omega_n) &=& 
G^{-1}({\rm i}\omega_n)-\Sigma({\rm i}\omega_n),
\eeqa
where ${\bf I}$ denotes a $2\times2$ unit matrix. The functional dependence of $\Sigma({\rm i}\omega_n)[G_0,U]$, defined on the discrete set of Matsubara frequencies $\omega_n=(2n+1)\pi T$, on $G_0$ and $U$ is determined by an auxiliary Anderson impurity problem, for a solution of which we employ the continuous-time quantum Monte-Carlo (CT-QMC) algorithm.\cite{Werner06}
For quantities obtained in the imaginary time domain, i.e.\ the spectral function and the dynamical susceptibility, the analytic continuation to the real-frequency axis is performed
using the maximum-entropy method.\cite{Jarrell96}
%%%%%%%%%%%%%%%%%%%%%%%%%%%%%%%%%%%%%
\section{Results and discussion}
\label{sec:3}
\subsection{Phase diagram} %
\label{sec:3a}
%%%%%%%%%%%
 \begin{figure}[t]
   \vspace*{2mm}
  \begin{center}
  \includegraphics[width=\hsize]{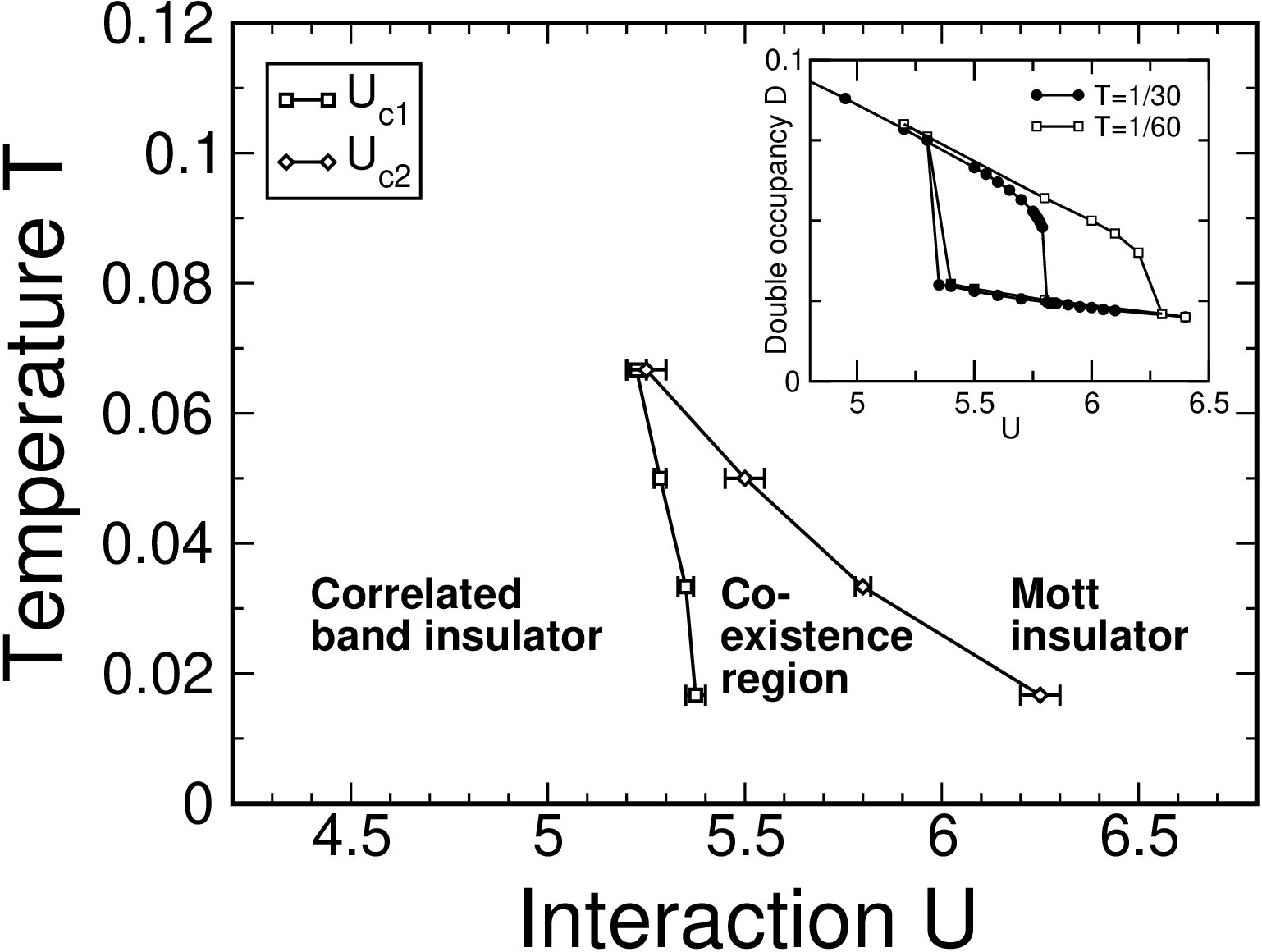} \\
  \end{center}
  \vspace* {2mm}
  \vspace* {-6mm}
      \caption{
      $T$-$U$ phase diagram at fixed $V=0.5$. The critical values of the
      interaction strength $U$ are determined from the double occupancy (see inset). Below a critical temperature both band and Mott insulating solutions of the DMFT equations are found depending on the initial guess for the self-energy. For temperatures above the critical end point of the coexistence region, there is a regime where the spectral function has a single peak at the Fermi energy accompanied by broad Hubbard bands (see Fig.\ \ref{fig:awt}).
      Inset: The double occupancy $D=\langle n_{\uparrow} n_{\downarrow}\rangle$ 
      as a function of $U$ for $T=1/30$ (circles) and $T=1/60$ (squares) 
      indicates the phase transition from the correlated BI (larger $D$) 
      to the MI (smaller $D$). 
      Both values of the double occupancy are
      shown in the region where two solutions are found
      in the DMFT self-consistency cycle. Note that all energies are given in units of the hopping integral $t=1$.
      \label{fig:phasediag}}
  \end{figure}
  %%%%%%%%%%
The non-interacting ground state ($U$=0) of our model 
is characterized by a gap in the spectral function of size $\Delta = 2V$ at any $V>0$.
In the $V=0$ limit we have two decoupled copies of a single-band Hubbard model with semi-circular density of states,
a problem which has been extensively studied within DMFT.\cite{Bulla01,Georges96}
It is well known that upon increasing the interaction strength $U$ at finite, low temperature the paramagnetic phase undergoes a
discontinuous transition from a metal to a MI with a hysteresis in the interval $U_{c_1}<U<U_{c_2}$.\cite{Bulla01} 
At high temperatures the hysteretic behavior is replaced by a continuous crossover.

In Fig.~\ref{fig:phasediag} we show the phase diagram in the $T$-$U$ plane obtained for $V=0.5$. 
At low temperatures two distinct phases exist separated by a discontinuous transition.
The phase boundaries were obtained by calculating
the double occupancy $D=\langle n_{\uparrow}n_{\downarrow}\rangle$, shown in the inset for two selected temperatures.
In a finite range of $U$ values we find two stable self-consistent solutions of the DMFT equations.\cite{comment:transition}
The two phases, adiabatically connected to the band insulator ($U=0$) and the Mott insulator ($U \rightarrow \infty$), respectively, both exhibit
a gap in the single-particle spectral function as discussed below. No signature of an intermittent metallic phase is found.
The calculated phase diagram resembles that of a single band Hubbard model including the existence of a critical end point of the
discontinuous phase transition, a weak $T$-dependence of $U_{c1}$ and a considerable increase of $U_{c2}$ upon lowering the temperature.\cite{Bulla01}
%%%%%%%%%%%%%%%%%%%%%%%%%%%
\subsection{Single-particle spectral function} %
\begin{figure}[t]
  \vspace* {2mm}
  \begin{center}
  \includegraphics[width=\hsize]{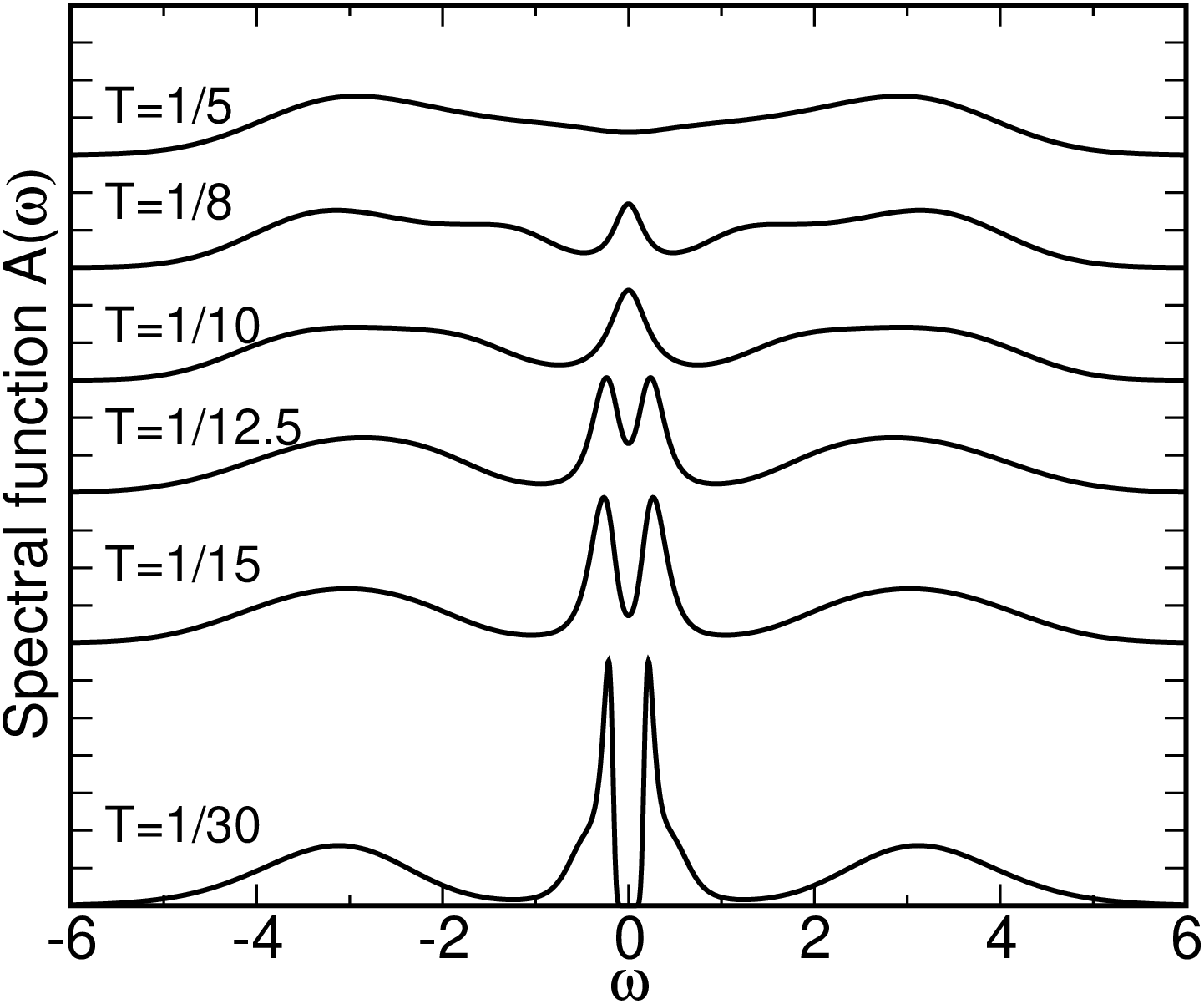} \\
  \end{center}
  \vspace* {-6mm}
      \caption{
      Local spectral function $A(\omega)$ at fixed hybridization $V=0.5$ and interaction strength $U=5$ for various
      values of the temperature $T$. Upon increasing $T$ the 
      correlated band insulator (split peak around $\omega=0$ plus Hubbard bands) shows a crossover to a Mott insulator (broad Hubbard bands and reduced spectral weight near $\omega=0$ at $T=1/5$). At intermediate temperatures ($T=1/8$ and $T=1/10$) there is a single peak at the Fermi energy accompanied by broad Hubbard bands. Note that the curves are shifted by a vertical offset for clarity.
      \label{fig:awt}}
  \end{figure}

\label{sec:3b}
%%%%%%%%%%%
  \begin{figure}[t]
  \vspace* {2mm}
  \begin{center}
  \includegraphics[width=\hsize]{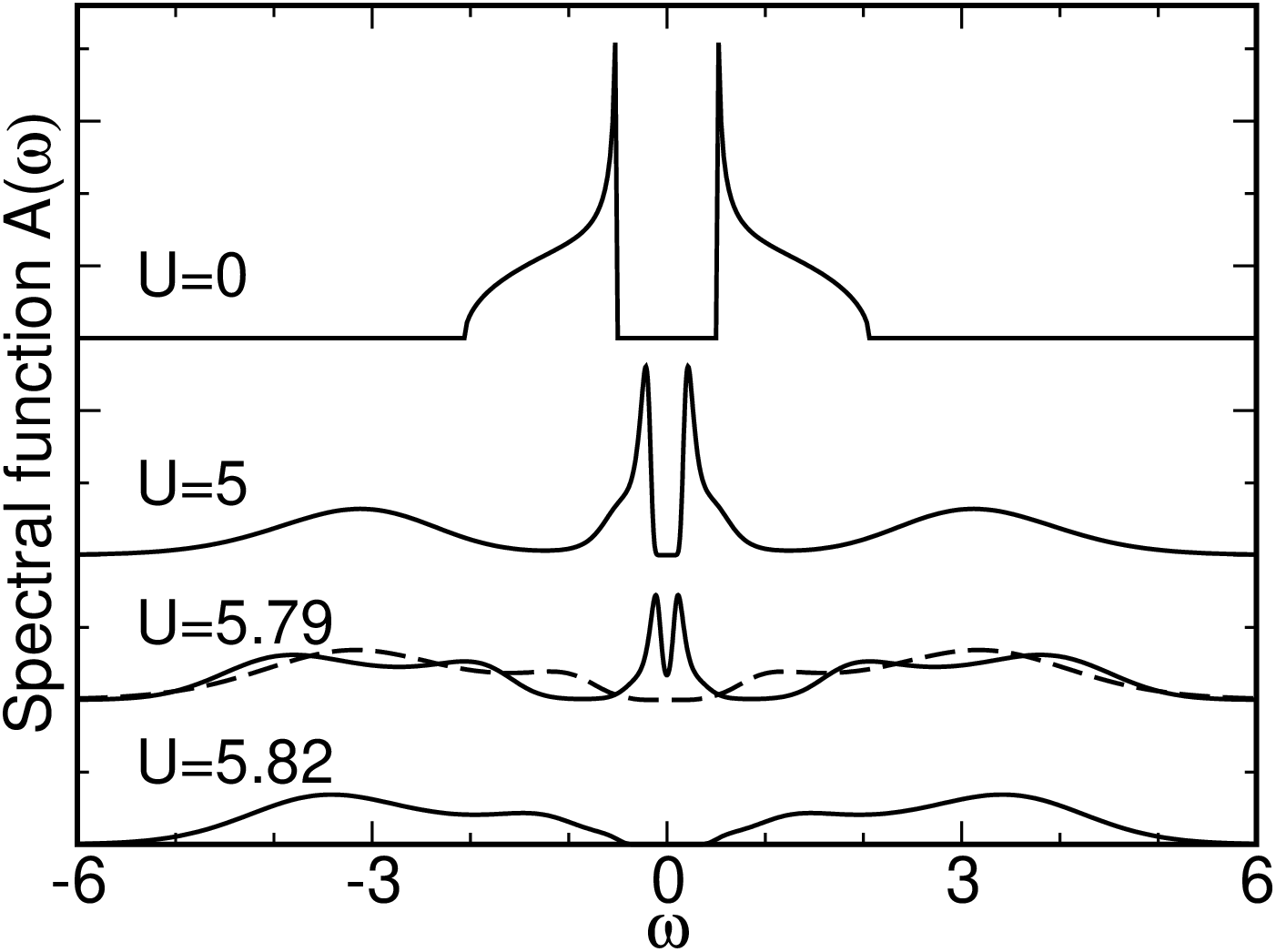} \\
  \end{center}
  \vspace* {-6mm}
      \caption{
      Local spectral function $A(\omega)$ at fixed
      interlayer coupling $V=0.5$ and temperature $T=1/30$ for various
      values of the interaction strength $U$. The  
      non-interacting density of states ($U=0$) has a charge gap
      $\Delta_c=2V=1$. The gap in the correlated band insulator shrinks with increasing $U$ until a discontinuous
      transition to a Mott insulator occurs with a      hysteresis
      region $5.35 < U < 5.82$. For $U=5.79$
      both   the      band (solid line) and Mott insulating (dashed line) solutions are
      displayed. All energies are given in units of the hopping integral $t=1$.
      \label{fig:awu}}
  \end{figure}
The evolutions of the spectral function $A(\omega)=- \text{Im}\; G(\omega+{\rm i}0^+)/\pi$ along a horizontal $T=1/30$ and a vertical $U=5$ line in the phase
diagram of Fig.~\ref{fig:phasediag} are shown in Figs.~\ref{fig:awu} and \ref{fig:awt}, respectively.
Remarkably, starting from $U=0$ the gap in the spectral function shrinks with increasing interaction strength $U$. At the same time the incoherent Hubbard bands
evolve and spectral weight is transferred to them. The gap is well distinguishable throughout the entire interaction range 
except for a small region in the vicinity of  $U_{c_2}$ where thermal broadening smears the strongly renormalized spectral features. 
 %%%%%%%%%%%%%
  
At low $T$ the spectral function in the band insulator phase close to the transition region consists of well distinguishable low-energy quasiparticle bands, separated by 
the hybridization gap, and incoherent Hubbard bands similar to the single-band Hubbard model.
Increasing the temperature the spectral gap is filled in while the quasiparticle bands lose their spectral weight. At $T=1/10$ the dip at chemical potential has vanished completely and 
a single peak remains. This peak smoothly disappears upon further increasing the temperature and at $T=1/5$ only two broad Hubbard bands remain in 
the spectrum reminiscent of the Mott insulator above the critical temperature of the metal-insulator transition.\cite{Bulla01}
In Ref.~\onlinecite{Kunes08} it was shown that this insulator--to--bad-metal crossover is reflected also in the {\it dc} and {\it ac} conductivity and
is accompanied by a substantial increase of the spin susceptibility which follows the Curie-Weiss law at high $T$.
%Also the double occupancy decreases considerably from $D=0.086$ at $T=1/30$ to $D=0.037$ at $T=1/5$. 
%In contrast, at weaker interaction strength ($U=3$, not shown) increasing the temperature from $T=1/30$ to $T=1/5$ 
%does not wipe out the split peak characteristic of the band insulator, and the double occupancy is much less 
%reduced  by the increasing temperature (from $D=0.153$ at $T=1/30$ to $D=0.136$ at $T=1/5$). 
%Remarkably a similar peculiar temperature dependence of the spectral density as shown in Fig.~\ref{fig:awt} was 
%found in the FeSi model study \cite{Kunes08}, where the filling of in-gap states and the corresponding decrease 
%in the resistivity at higher temperatures could not be explained by Fermi function effects alone.
%%%%%%%%%%%%%%%%%%%%%%%%%%%%%%%%%%%%%%%%%%%%
\subsection{Gap renormalization and self-energy}
\label{sec:3c} 
The spectral densities of Fig.~\ref{fig:awu} reveal a reduction of the charge gap in the CI phase with increasing $U$. In this section
we analyze this behavior which is quantified in Fig.~\ref{fig:gap} showing the charge gap $\Delta_c(U)$ deduced from the
spectral densities. In the following we derive the gap renormalization in two different ways from second order perturbation 
theory. Our derivation closely follows the approach of Ref.~\onlinecite{Ruhl06}.
%%%%%%%%%%%%%%%%%%%%%%%%%%%
\subsubsection{Renormalization of the charge gap deduced from the total energy}
The charge gap $\Delta_c$ in the state with $N$ particles is defined as
\begin{equation}
\label{eq:E_n}
\Delta_c=(E(N+1)-E(N))+(E(N-1)-E(N)),
\end{equation}
where $E(N)$ is the ground-state energy of the system with $N$ particles.
The first correction to (\ref{eq:E_n}) from a perturbative expansion in $U$
comes from the 2nd order diagram. 
Using the factorization of the joint density of states in the limit of infinite dimensions the
second order correction to the ground-state energy can be written as
\begin{equation}
\label{eq:e2}
E^{(2)}=-\frac{U^2}{L^3}\sum_{p_1,p_2,p_3,p_4}\frac{(1-n_{p_1\uparrow})n_{p_2\uparrow}(1-n_{p_3\downarrow})n_{p_4\downarrow}}{\epsilon_{p_1}-\epsilon_{p_2}+
\epsilon_{p_3}-\epsilon_{p_4}},
\end{equation}
where $n_{p_i}$ is the occupation number and $\epsilon_{p_i}$ is the energy of the single-particle state with index $p_i$. The non-interacting $N$-particle
ground state is a band insulator with all states with energies $\epsilon<-V$ filled. The non-interacting ($N+1$)-particle ground state is obtained by filling
the lowest energy state of the empty conduction band $\epsilon_{p_0}=V$ (we choose spin $\downarrow$ from the two possibilities). Keeping only the terms that do not vanish in the
thermodynamic limit and using particle-hole symmetry, which requires that $\Delta_c=2(E(N+1)-E(N))$, we obtain the second order correction to the charge gap
\beq
\begin{split}
\label{eq:diff}
&\Delta_c^{(2)}=-2\frac{U^2}{L^3}
\sum_{p_1,p_2,p}
(1-n_{p_{1}\uparrow})n_{p_{2}\uparrow} \\
&\times
\left[
\frac{1-n_{p\downarrow}}{\epsilon_{p_1}-\epsilon_{p_2}+\epsilon_{p}-V}
-\frac{n_{p\downarrow}}{\epsilon_{p_1}-\epsilon_{p_2}+V-\epsilon_{p}}
\right].
\end{split}
\eeq
Note that while $E^{(2)}$ is an extensive quantity, the difference in (\ref{eq:diff}) remains finite when $L\rightarrow\infty$. Introducing
the single-particle density of states $D$ and its Laplace transform $F$
\begin{equation}
\label{eq:DOS}
D(\epsilon)=\frac{1}{L}\sum_p\delta(\epsilon-\epsilon_p),\quad
F(\lambda)=\int_0^{\infty} \text{d}\epsilon\; e^{-\lambda\epsilon}D(\epsilon),
\end{equation}
(\ref{eq:diff}) can be rewritten as \cite{Ruhl06}
\beq
\label{eq:diff_2}
\begin{split}
&\Delta_c^{(2)}=-2U^2\int \text{d}\epsilon_1\; \text{d}\epsilon_2\; \text{d}\epsilon_3\;
D(\epsilon_1)D(\epsilon_2)D(\epsilon_3)\\
&\times
(1-n_1)n_2
\left[
\frac{1-n_3}{\epsilon_1-\epsilon_2+\epsilon_3-V}-\frac{n_3}{\epsilon_1-\epsilon_2+V-\epsilon_3}
\right]\\
&=-4U^2\int_0^{\infty} \text{d}\lambda\; \operatorname{\sinh}({\lambda V})F^3(\lambda),
\end{split}
\eeq
where the fixed spin index has been dropped for simplicity. Introducing the bare gap $\Delta_c^0=2V$ we can write the renormalized charge gap as
\begin{equation}
\label{eq:eg_ren0}
\Delta_c=\Delta_c^0-4U^2\int_0^{\infty} \text{d}\lambda\; \operatorname{\sinh}\left(\lambda \frac{\Delta_c^0}{2}\right)F^3(\lambda).
\end{equation}
This expression shows that the reduction of the charge gap does not depend on the details of $D(\epsilon)$, but rather on its overall characteristics
such as the total bandwidth. If $\Delta_c^0$ is small compared to the total bandwidth we can linearize
the expression (\ref{eq:eg_ren0}) to obtain
\begin{equation}
\label{eq:eg_ren}
\Delta_c=\Delta_c^0\left(1-2U^2\int_0^{\infty} \text{d}\lambda\; F^3(\lambda)\lambda\right).
\end{equation}
In this limit the gap acquires a simple multiplicative renormalization, which is closely related to  
the quasiparticle mass renormalization as shown below. The perturbative results are compared with the
numerical data in Fig.~\ref{fig:gap}.

We close this section with two remarks. First, equation (\ref{eq:e2}) relies on the equality
of the local and the total density of states, which is where the concept of a covalent insulator enters the
algebra. Second, the physical origin of the reduction of the gap is best seen in (\ref{eq:diff}). Adding a single
electron to the insulating states blocks scattering processes with a contribution of the order $-1/4V$ but
adds the same number of processes contributing $-1/2V$ and therefore leads to an overall gain in the 
correlation energy in the ($N+1$)-particle state and thus a reduction of the charge gap from its non-interacting value.
%%%%%%%%%%%%%%%%%%%%%%%%%%%
\subsubsection{Renormalization of the charge gap deduced from the self-energy}
%%%%%%%%%%%
  \begin{figure}[t]
  \begin{center}
  \includegraphics[width=\hsize]{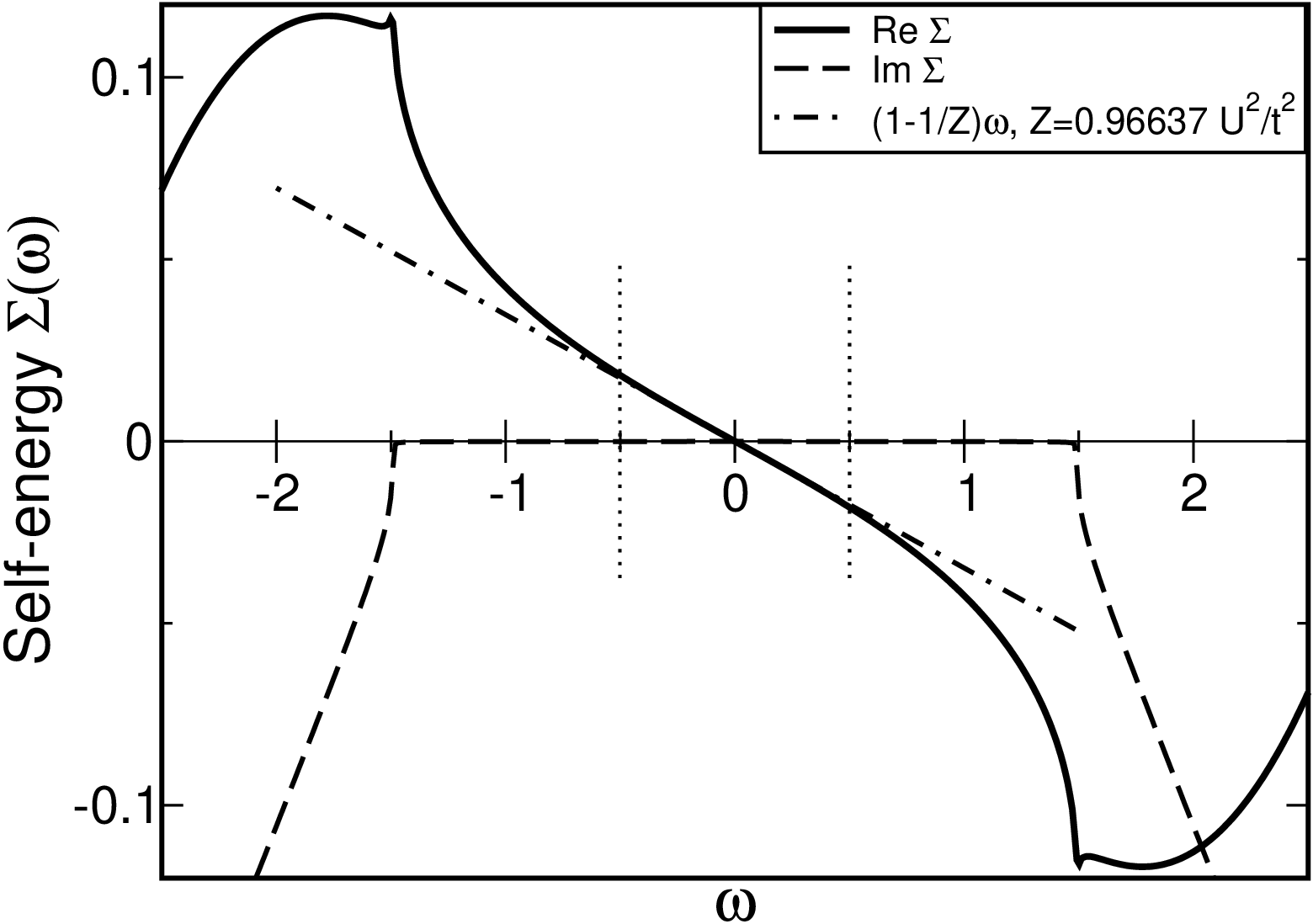} \\
  \end{center}
  \vspace* {-3mm}
      \caption{
      The self-energy within second-order weak-coupling perturbation theory at zero temperature (see Eq.\ (\ref{eq:sigma0})) for $V=0.5$. The real part (solid line) is linear around $\omega=0$; the slope (indicated by the dot-dashed line) determines the $Z$ factor. The imaginary part (dashed line) is gapped for $|\omega|<3V$. The dotted vertical lines indicate the bare gap $\Delta_c^0=1$. The scale of the vertical axis is $U^2/t$.
\label{fig:sigma-perturb}}
\end{figure}
  
An alternative derivation of the gap renormalization can be obtained from the perturbative calculation
of the self-energy in the insulating ground state. Using the factorization of the joint density of states
as in the previous section, the second order contribution to the self-energy is written as
\beq
\label{eq:sigma0}
\begin{split}
&\Sigma(\omega)=U^2\int \text{d}\epsilon_1\; \text{d}\epsilon_2\; \text{d}\epsilon_3\;
D(\epsilon_1)D(\epsilon_2)D(\epsilon_3)\\
&\times
\frac{(1-n_1)n_2n_3
+n_1(1-n_2)(1-n_3)}{\epsilon_1-\epsilon_2+\omega-\epsilon_3},
\end{split}
\eeq
where the spin indices were dropped as in the previous section.
For $-3V<\omega<3V$ the denominator of the integrand remains negative throughout the entire integration range,
the self-energy is therefore real and can be expressed using the Laplace transform (\ref{eq:DOS}) of the non-interacting
density of states
\begin{equation}
\label{eq:sigma}
\Sigma(\omega)=-2U^2\int_0^{\infty} \text{d}\lambda\; \operatorname{\sinh}(\lambda \omega)F^3(\lambda).
\end{equation}
The renormalization of the band gap is obtained by searching for the pole $\Omega$ of the renormalized propagator of the lowest
unoccupied state $\epsilon_{p_0}=V$:
\begin{equation}
\Omega-V-\Sigma(\Omega)=0.
\end{equation}
In the small $U$ limit we can replace $\Sigma(\Omega)$ by $\Sigma(V)$ and with $\Delta_c=2\Omega$ we recover
equation (\ref{eq:eg_ren0}).
As in the previous section we can linearize (\ref{eq:sigma}) for small $V$ in the interval
$|\omega|\lesssim V$, 
which leads to
\beq
\begin{split}
&\Delta_c=Z\Delta_c^0, \\
\text{where}\quad
 & Z = \left(1-
      \left. \frac{\partial {\rm Re}\; \Sigma(\omega)}{\partial\omega}
      \right\vert_{\omega=0}\right)^{-1}.
\end{split}
\label{eq:qpweight}
\eeq
For $|\omega|>3V$ the second order self-energy acquires a finite imaginary part and expression
(\ref{eq:sigma}) is not applicable.
The second order self-energy obtained by numerical integration of (\ref{eq:sigma0})
is shown in Fig.~\ref{fig:sigma-perturb}.
  \begin{figure}[t]
  \begin{center}
  \includegraphics[width=\hsize]{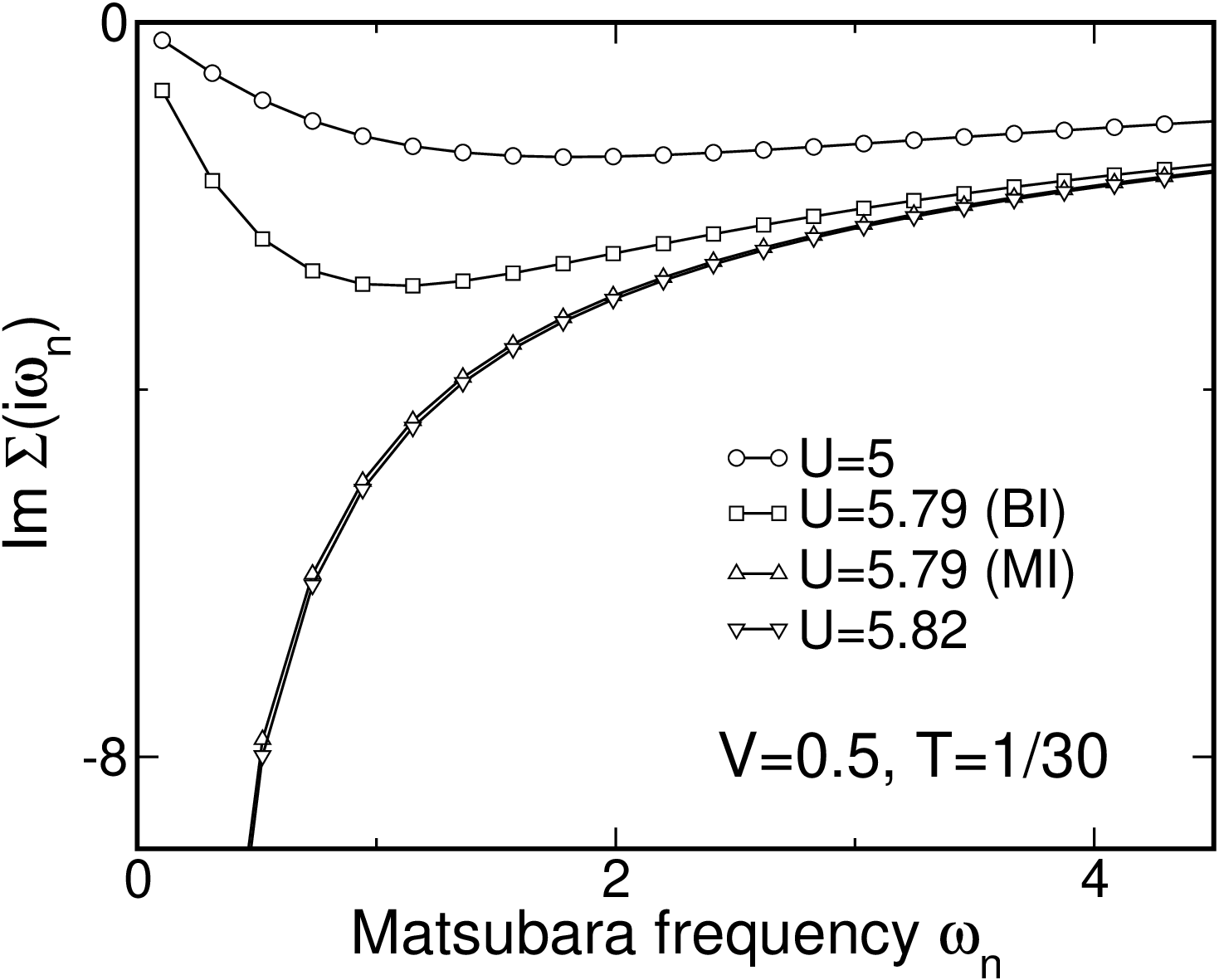} \\
  \end{center}
  \vspace* {-6mm}
      \caption{
      Imaginary part of the self-energy as a function of Matsubara
      frequencies $\omega_n$ for three values of the interaction strength
      $U$. In the correlated BI the self-energy has a negative slope
      at small $\omega_n$. In the coexistence region
      ($U=5.79$) both solutions are displayed.     
      \label{fig:sigma}}
  \end{figure}

The linear behavior of ${\rm Re}\; \Sigma(\omega)$ around the chemical potential is reminiscent of a Fermi liquid. It is not limited
to second order perturbation theory, but is a general consequence of a sufficiently fast vanishing of ${\rm Im}\; \Sigma(\omega)$
in the vicinity of the chemical potential and the Kramers-Kronig relations. While in Fermi liquids ${\rm Im}\; \Sigma(\omega)\sim\omega^2$,
the existence of a gap in the spectral function of the covalent insulator, as found in the numerical simulations, 
and the absence of a pole in $\Sigma(\omega)$ inside the gap, imply ${\rm Im}\; \Sigma(\omega)=0$ throughout the entire gap. As a result,
${\rm Re}\; \Sigma(\omega)$ of the CI phase closely resembles the self-energy of a Fermi liquid. This similarity 
is made evident in Fig.~\ref{fig:sigma} where we show the self-energy on the discrete set of Matsubara frequencies: It is barely distinguishable from 
the analogous plot obtained for the single-band Hubbard model with comparable parameters.\cite{Bulla01}
%%%%%%%%%%%%%%%%%%%%%%%%%%%
\subsection{Spin excitations in the band insulator} %
\label{sec:3d}
Typically a band insulator is characterized by identical gaps for charge and spin excitations. So far we have identified the insulator below the critical interaction strength as a ``correlated band insulator'' due to the fact that it is adiabatically connected to the band insulator at $U=0$. Naturally the question arises whether spin and charge gaps remain indeed equal in the presence of correlations. To answer this question and to characterize more precisely the correlated band insulator we evaluate the dynamical spin susceptibility and compare spin and charge gaps.

While the spectral function $A(\omega)$ is
gapped in both band and Mott insulators (see Fig.~\ref{fig:awu}), the spin gap is finite
in the BI and zero in the MI. The spin excitation spectrum is reflected in the local dynamical spin susceptibility $\chi_s(\omega)$, which
is calculated by a QMC measurement of the imaginary time correlation
function $\chi_s(\tau) = \langle S_z(\tau) S_z(0)\rangle$ and the
analytic continuation of its Matsubara transform to real frequencies. Here we use the maximum
entropy method \cite{Jarrell96} for the bosonic kernel according to 
\beq
\chi_s(\tau)=\frac{1}{\pi} \int \text{d}\omega \frac{e^{-\tau\omega}}{1-e^{-\beta\omega}} \text{Im}\;\chi_s(\omega).
\eeq

\begin{figure}[t]
  \begin{center}
  \includegraphics*[width=\hsize]{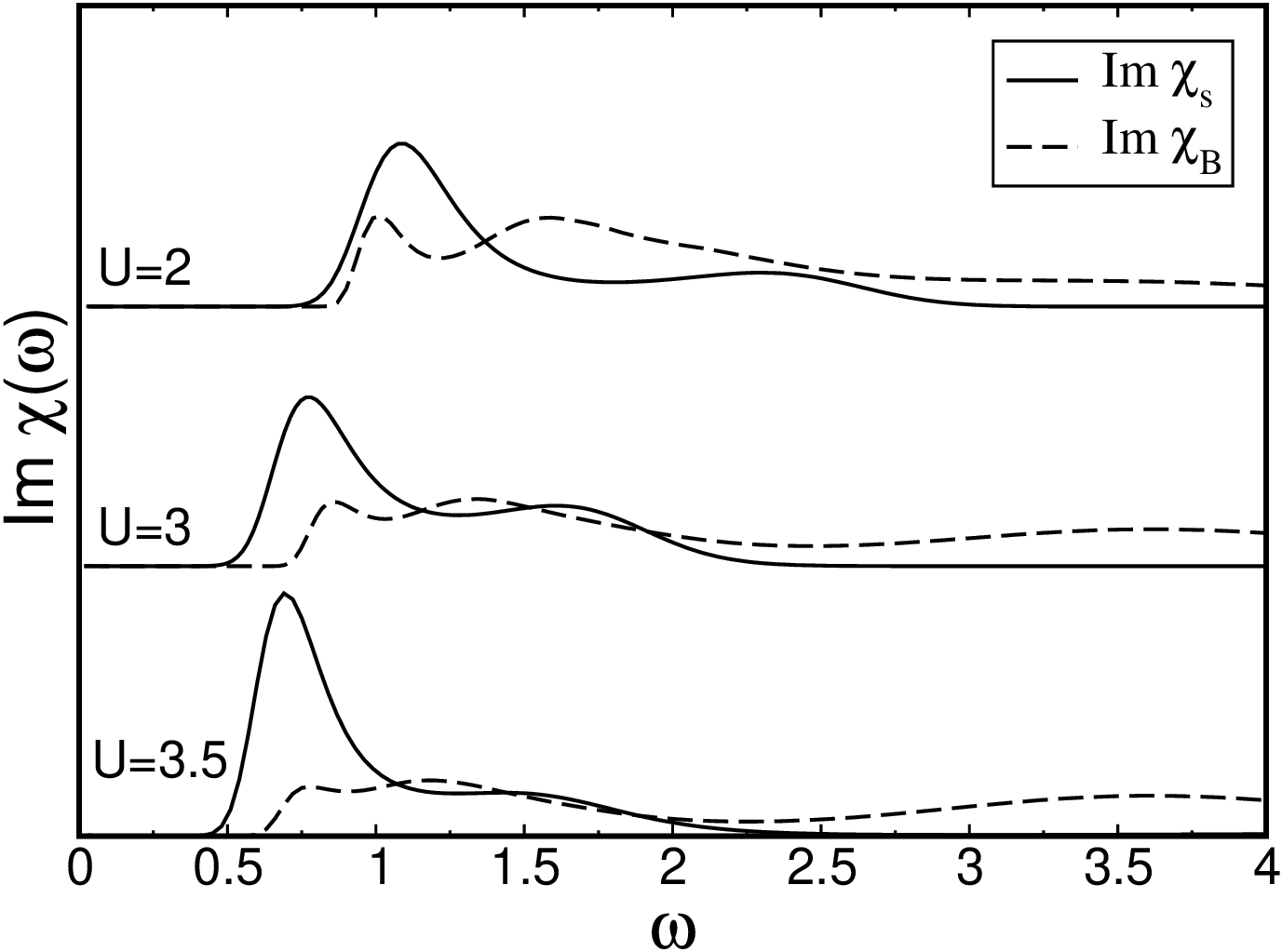} \\
  \end{center}
  \vspace* {2mm}
  \vspace* {-6mm}
      \caption{
      Local dynamical susceptibilities in the correlated BI. At
      $V=0.5$ and temperature $T=1/30$ the imaginary parts of the
      spin susceptibility $\chi_s$ (solid lines) and the bubble
      diagram $\chi_{B}$ (dashed lines) are shown for $U\in\{2, 3, 3.5\}$.
      The susceptibilities are obtained from the QMC data by analytic continuation. The bubble diagram is calculated from the convolution of the fully-dressed Green functions  (see Eq.~(\ref{eq:bubble})). Thus the gap in ${\rm Im}\;\chi_{B}(\omega)$ corresponds to the single-particle energy gap in $A(\omega)$, which is apparently larger  than the spin gap observed in  ${\rm Im}\;\chi_s(\omega)$ in the correlated insulator.
      \label{fig:susc}}
  \end{figure}

In Fig.~\ref{fig:susc}
the imaginary part of the spin susceptibility is shown
on the real-frequency axis for $V=0.5$, $T=1/30$, and $U\in\{2, 3,
3.5\}$. Similar to the charge gap, the spin gap also shrinks with increasing interaction strength. For a quantitative comparison of correlation effects on spin and charge excitations in the correlated band insulator Fig.~\ref{fig:susc} also
shows the bubble diagram calculated from the convolution of the fully
dressed local Green functions,
\beq
{\rm Im}\;\chi_{B}(\omega)
=
\pi \int \text{d}\epsilon \frac{A({\epsilon}) A({\omega-\epsilon})\left(1-e^{-\beta\omega}\right)}{(1+e^{-\beta\epsilon})(1+e^{-\beta(\omega-\epsilon)})}.
\label{eq:bubble}
\eeq
In the non-interacting limit (not shown) ${\rm Im}\;\chi_{B}(\omega)$ and ${\rm Im}\;\chi_s(\omega)$ coincide and exhibit an energy gap of size $2V=1$. In contrast at finite $U$ the curves differ considerably from each other. In the spin susceptibility a prominent peak develops at energies lower than the charge gap. Furthermore spectral weight is suppressed at higher energies which correspond to excitations from the split central peak to the Hubbard bands. The spin susceptibility is thus both qualitatively and quantitatively more strongly influenced by correlations in comparison to the bubble diagram, which contains correlation effects via the renormalized propagator only. The effect of correlations on the band insulator therefore goes beyond the energy gap renormalization discussed in Sec.\ \ref{sec:3c}.

\begin{figure}[t]
  \begin{center}
  \includegraphics[width=\hsize]{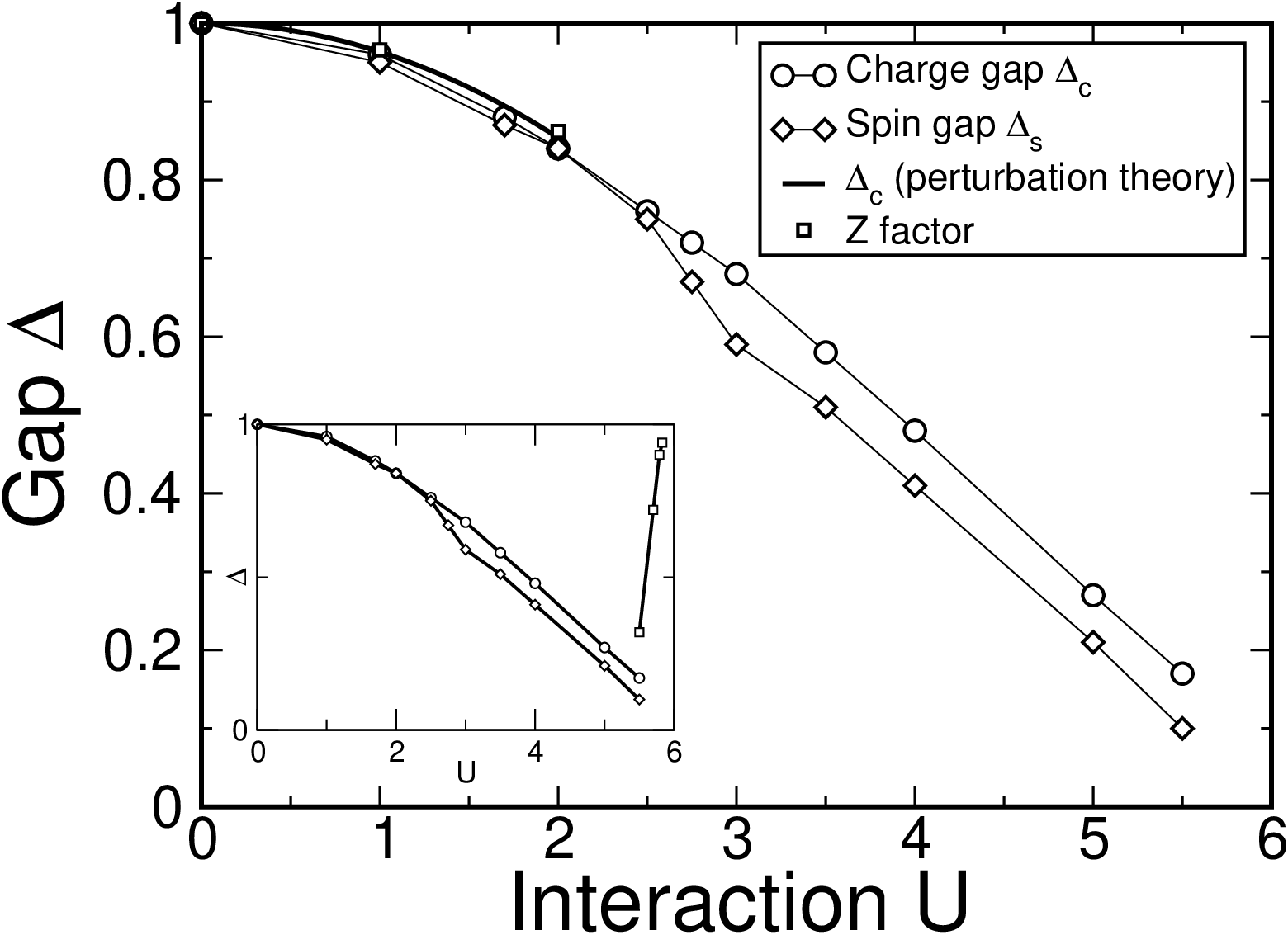} \\
  \end{center}
  \vspace* {2mm}
  \vspace* {-6mm}
      \caption{
      Spin and charge gaps in the correlated BI as a function of $U$
      for $V=0.5$ as determined from the spectral function and the spin
      susceptibility at $T=1/30$, respectively.\cite{gapcomment} The thick solid line
      shows the charge gap obtained from second order perturbation
      theory
      (Eq.~(\ref{eq:eg_ren0})).
      The squares represent the $Z$ factor (Eq.~(\ref{eq:qpweight})).
      Inset: Discontinuous change of spin and charge gaps at the BI to MI transition. In the Mott insulator $\Delta_s=0$.
      \label{fig:gap}}
  \end{figure}

The comparison of the spin susceptibility and the bubble diagram in Fig.~\ref{fig:susc} shows that charge and spin gap do not coincide in the correlated BI. For a quantitative analysis we extract the gap values by a linear
extrapolation to the frequency axis using the slope at the inflexion point of the
spectral function and the imaginary part of the spin susceptibility for the charge and the spin gap, respectively. Fig.~\ref{fig:gap}
shows the evolution of charge and spin gaps with increasing interaction strength in the correlated band insulator. As discussed in Sec.\ \ref{sec:3c} the energy gap renormalization at moderate coupling strengths is well described by the $Z$ factor (Eq.~(\ref{eq:qpweight})) extracted from the slope of the self-energy at low frequencies. For $U \gtrsim 2.5$ the spin gap is smaller than the charge gap, and $\Delta_c-\Delta_s$ remains almost constant with increasing the interaction strength further. Therefore, in contrast to the non-interacting limit, the description of the band insulator as an insulator with identical energy gaps for both charge and spin excitations no longer holds once the interaction is strong enough. 

It is worthwhile to point out that different spin and charge gaps were also obtained in the half-filled one-dimensional ionic Hubbard model in the weakly correlated regime, when the ionic potential is smaller than the onsite interaction $U$.\cite{Kampf03,Noack04} As in the covalent insulator, $\Delta_c$ and $\Delta_s$ decrease with increasing $U$. However, the transition to the Mott insulator proceeds via an intermediate insulating phase with bond order and a staggered modulation of the kinetic energy on neighboring bonds.

Hints for a possible difference of spin and charge gaps as obtained in the above discussed correlated band insulator phase exist also from experiments on selected insulating materials. In their study of FeSi Schlesinger \emph{et al.} \cite{Schlesinger97} pointed out the
possible difference between spin and charge gaps in correlated insulators based on evidence for a larger charge gap in the Kondo insulator Ce$_3$Bi$_4$Pt$_3$.\cite{gaps} While the single-particle charge gap can be measured with $meV$ accuracy using photoemission
spectroscopy, the determination of the spin gap requires accurate
susceptibility measurements to temperatures much lower than the gap
energy. The relative gap difference for FeSi, e.g., is expected to be less than 30\ \% and could so far not be resolved from existing measurements.\cite{Schlesinger97}  
%%%%%%%%%%%%%%%%%%%%%%%%%%%
\section{Conclusion} %
\label{sec:conclusion}
We have studied correlation effects in a covalent band insulator using dynamical mean-field theory. In the absence of correlations a band insulator is characterized by its band gap. A local Coulomb repulsion renormalizes the energy gap, which surprisingly shrinks when the interaction strength is increased. In second-order perturbation theory the gap shrinking can be traced to enhanced low-energy scattering phase space in the conduction band. By analogy to the quasiparticle weight in Fermi-liquid theory a renormalization factor $Z$ can also be introduced in interacting insulators based on the low-frequency behavior of the self-energy. The $Z$ factor in the insulator determines the energy gap renormalization. The simple one-gap picture of the band insulator breaks down for sufficiently large interaction strengths. In the correlated band insulator the spin gap is smaller than the charge gap. A discontinuous transition from the band to the Mott insulator occurs upon increasing the Coulomb repulsion at low but finite temperature. Close to the insulator-insulator transition the increase of temperature in the correlated band insulator with a split central peak and pronounced Hubbard bands leads to a crossover into a high-temperature Mott insulator phase with broad Hubbard bands. 

The correlation-driven reduction of the energy gap is rare among the known materials. Nevertheless, this mechanism provides a natural explanation of the uncommon gap overestimation in band-structure calculations of systems like FeSi.\cite{Menzel09}
Tracing the difference between spin and charge gaps in the correlated band insulator to its physical origin remains a task for future work. 
%%%%%%%%%%%%%%%%%%%%%%%%%%%
\vspace{4mm}

  \centerline{\bf{Acknowledgement}}
  
  We thank Dieter Vollhardt and Vladimir Anisimov for insightful discussions. M.S.\ acknowledges support
  by Studienstiftung des Deutschen Volkes. This work was supported in
  part by the SFB 484 of the DFG. The allocation of CPU time on
  the supercomputer HLRB II by the Leibniz-Rechenzentrum in Munich
  is gratefully acknowledged. The calculations were performed using
  the ALPS library.\cite{ALPS}

  %%%%%%%%%%%%%%%%%%%%%%%%%%%%%%%%%%%%%%%%%%%%%%%%%%%%%%%%%%%%% 

%  \vspace*{-6mm}


\begin{thebibliography}{99}
\bibitem{comment:transition}
    For the determination of the critical interaction strength of the BI-MI transition it is necessary to determine and compare the free energy of both coexisting solutions.
\bibitem{gapcomment}
    Note that spin and charge gaps appear identical for $U=2$. In Fig.~\ref{fig:susc}, however, the gaps in the spin susceptibility and the bubble diagram already differ from each other at $U=2$. This seeming quantitative discrepancy is due to the fact that we determine the charge gap directly from the spectral function and not from the bubble diagram. In fact the charge gap is obtained slightly larger if it is extracted from the bubble diagram.
\end{thebibliography}
\end{document}